\documentclass[useAMS,onecolumn]{mn2e}
\usepackage{epsfig}
\title[Binary destruction in ellipticals]{Destruction of wide binary stars in low mass elliptical galaxies: implications for initial mass function estimates}

\author[Maccarone et al.]{Thomas J. Maccarone\\ Department of Physics, Texas Tech University, Box 41051, Lubbock TX 79409-1051, USA\\}

\begin{document}
\def\ltsim{\mathrel{\rlap{\lower 3pt\hbox{$\sim$}}
        \raise 2.0pt\hbox{$<$}}}
\def\gtsim{\mathrel{\rlap{\lower 3pt\hbox{$\sim$}}
        \raise 2.0pt\hbox{$>$}}}

\date{}

\pagerange{\pageref{firstpage}--\pageref{lastpage}} \pubyear{}

\maketitle

\label{firstpage}

\begin{abstract}
We discuss the effects of destruction of wide binaries in the nuclei
of the lower mass giant elliptical galaxies.  We show that the numbers
of barium stars and extrinsic S stars should be dramatically reduced
in these galaxies compared to what is seen in the largest elliptical
galaxies.  Given that the extrinsic S stars show strong Wing-Ford band
and Na I D absorption, we argue that the recent claims of different
initial mass functions from the most massive elliptical galaxies
versus lower mass ellipticals may be the result of extrinsic S stars,
rather than bottom-heavy initial mass function.

\end{abstract}

\begin{keywords}
stars:binaries -- stars:chemically peculiar -- galaxies:stellar content -- supernovae:general
\end{keywords}

\section{Introduction}

Despite the fact that most stars are members of binary systems, binary stellar
evolution is usually neglected in modelling of the spectral energy
distributions of galaxies (although see e.g. Han et al. 1995).  This decision
is made as a basic simplication, in part because it has not, to date, been
clear how binary evolution would affect most of what is seen in optical light,
and in part because binary population synthesis is extremely complicated, with
large numbers of parameters which are not particularly well-constrained by
observations (see e.g. Belczynski et al. 2008 for a discussion of a particular
binary population synthesis code).  Furthermore, while we will show evidence
to the contrary in this paper, one might initially expect that the effects of
binary evolution would not vary much from galaxy to galaxy.

Heggie's Law (1975) states that hard binaries get harder while soft binaries
get softer -- i.e. binaries whose binding energy is larger than the mean
kinetic energy of single stars in their local neighborhood will tend to become
closer with time, while binaries with binding energies less than the local
mean stellar kinetic energy will tend to become wider with time until they are
eventually dissolved into two single stars.  The consequences of Heggie's Law
are generally well-appreciated, if not fully understood, in the context of
globular clusters.  For example, the hardening of binaries in globular
clusters supplies kinetic energy to the single stars in the clusters, holding
up the collapses of star clusters in a manner somewhat analogous to the manner
in which nuclear fusion holds up the collapses of stars (see e.g. Sugimoto \&
Bettweiser 1983; Fregeau 2008).

In the context of field populations of galaxies, it has been shown
that the absence of wide binaries in the Galactic halo can be taken as
evidence against massive compact halo objects (i.e. MACHOs) supplying
the bulk of the dark matter in the Milky Way (Yoo et al. 2004).  There
has been relatively little appreciation, however, of how removing long
period binaries from a stellar population affects integrated stellar
light.  Traditionally, in fact, stellar population synthesis models
for understanding galaxy evolution have ignored binaries almost
entirely, except with respect to binary models for producing the
ultraviolet upturn in elliptical galaxies (e.g. Han et al. 2002; Han
et al. 2007), and, of course population synthesis calculations aimed
at unequivocally binary populations like X-ray binaries and double
neutron stars (e.g. Belczynski et al. 2008).  In this Letter, I will
show that the cutoff period varies considerably through different
classes of stellar systems, and that this difference affects whether
Roche lobe overflowing red giants will be present in different classes
of galaxies.  I will show further that these binary systems may then
have profound implications for the observational appearance of
different classes of galaxies.

\section{Binary dissolution timescales}

Binney \& Tremaine (2009) give the dissolution timescale of a binary as:
\begin{equation}
t_d = 15 {\rm Gyr} \left(\frac{K_{diff}}{0.002}\right)\left(\frac{\sigma_{rel}}{40 {\rm km/sec}}\right)\left(\frac{M_b}{2 M_\odot}\right)\left(\frac{M_p}{M_\odot}\right)^{-1}\left(\frac{0.05 {\rm pc^{-3}}}{n}\right)\left(\frac{10^4 {\rm AU}}{a}\right),
\end{equation}
where $K_{diff} = \frac{0.022}{{\rm ln}\Lambda}$ (and $\Lambda =
\frac{\sigma_{rel}a}{GM_p}$), so that $K_{diff} \approx 0.002$ for
nearly all galaxies; $\sigma_{rel}$ is the velocity dispersion of the
scattering stars, $M_p$ is the mass of the star perturbing the binary,
$M_b$ is the mass of the binary, $n$ is the number density of stars in
the local region, and $a$ is the orbital separation of the binary.

Now, let us consider two relatively extreme giant elliptical galaxies: M87,
the central galaxy in the Virgo Cluster, and NGC~4458, a small galaxy in the
Virgo Cluster.  For M87, the central velocity dispersion is about 400 km/sec,
and the central density is about 200 stars per cubic parsec (Gebhardt \&
Thomas 2009).  For NGC~4458, the central density is about 2800
$L_\odot$/pc$^3$ (Gebhardt et al. 1996), and the central velocity dispersion
in 85 km/sec (van Dokkum \& Conroy 2011).  In general, the fundamental plane
relation (Dressler et al. 1987) shows giant elliptical galaxies which fit on
the plane will become significantly denser and have significantly lower
velocity dispersions as their masses drop.  For a 10 Gyr old population in a
dynamical environment like that of M87, binary separations of up to about 20
AU will be possible in the core; for a galaxy like NGC~4458, binary
separations of about 0.4 AU will be possible.

\section{Binary evolution at around 1 AU separation}

Two important classes of objects have orbital periods of order 1 year.
They may be destroyed in the cores of the smallest giant elliptical
galaxies, but not in the cores of the largest galaxies.  They are
symbiotic stars, and barium stars/extrinsic S stars.

First, let us consider the case of symbiotic stars.  These are binary
systems which contain compact objects which accrete from a highly
evolved companion star -- either a red giant or an asymptotic giant
branch star.  While there are a few neutron star symbiotic binaries,
the vast majority of the symbiotic stars have white dwarf accretors
(Belczynski et al. 2000).  The orbital periods of the catalogued
symbiotic stars range from a little over 200 days to 5700 days
(Belczynski et al. 2000).  Even the shortest period symbiotics have
separations of about 0.9 AU (assuming $P=250$ days and a total system
mass of about $1.5M_\odot$).  One reason for particular interest in
the symbiotic stars is that a disproportionate fraction of recurrent
novae occur in symbiotic stars, as are a disproportionate fraction of
the steady supersoft sources (e.g. Sokoloski 2003).  It is not
generally considered that these objects represent a large fraction of
Type Ia supernovae (e.g. Schaefer 2014), but searching for a deficit
of central Type Ia's in dense galaxies would represent an additional
possible test.

The other class is a group of stars with enhancements in their
s-process element abundances -- barium stars, CH stars and extrinsic S
stars (sometimes called Tc-poor S stars).  Barium stars (see Bidelman
\& Keenan 1951 for the discovery of the class, and e.g. Warner 1965;
McClure 1985 for a working definition of the class) are G/K giants
which show exceptionally strong absorption lines from s-process
elements - especially barium and strontium.  CH stars (Keenan 1942)
represent the Population II analogs of the barium stars.  S stars in
general represent the class of cool stars rich in s-process elements
(Merrill 1922 -- although at the time of the establishment of the
category, they were called S stars with the letter S being chosen,
apparently, arbitrarily with the connection to the $s-$ process
coincidental).  It has become appreciated in recent years that {\it
  all} barium stars are found in wide binaries (i.e. with orbital
periods of at least 600 days -- Jorissen \& Mayor 1992) with white
dwarf companions, in accord with theoretical predictions (e.g. Iben \&
Renzini 1983).  This finding has led to a model for their production
in which a star donates most of its envelope to its companion star
after it evolves off the main sequence, so that the companion star's
new abundances become the abundances of the interior of the originally
more massive star (see e.g. McClure \& Woodsworth 1990; Han et
al. 2002).  Extrinsic S stars are the S stars thought to form through
binary stellar evolution, as descendants of the barium stars.  There
also exist intrinsic S stars, which are thought to form as s-process
elements are raised to the surface in the third dredge-up in the
evolution of a moderately massive single stars.  The most common means
of distinguishing between the two is by searching for lines from Tc,
an s-process element with a half-life of order a million years --
similar to the lifetimes of the intrinsic S stars, but much shorter
than the lifetimes of the extrinsic S stars.

\subsection{Implications for supernova progenitors}

The implications for our understanding of Type Ia supernova
progenitors are quite clear.  If symbiotic stars dominate the
progenitors of Type Ia supernovae in old stellar populations, then
there should be a strong deficit of such objects seen from the centers
of small elliptical galaxies.  The Type Ia supernova rates on the
outskirts of these galaxies, where the stellar density has dropped,
should not be affected.  Searching the centers of the highest surface
brightness galaxies is not easy, but may pay large dividends.  On a
related note, any elements predominantly produced in recurrent novae
may be preferentially more abundant in the largest giant ellipticals
than in smaller giant ellipticals.

\subsection{S stars: implications for understanding integrated light spectroscopy from galaxies}

The use of certain near-infrared spectral features to constrain the
initial mass function of stars has been suggested for quite some time.
Whitford (1977) suggested that the suggested that the FeH band at 9916
\AA (discovered by Wing \& Ford in 1969, and hence often called the
Wing-Ford band) could be used to estimate the amount of light from
dwarf stars in old stellar populations, but did not detect the band in
a sample of seven galaxies he observed.  Hardy \& Couture (1988) did
detect the band, but noted that the presence of a nearby TiO feature
would often complicate the interpretation of such measurements.  Using
more sophisticated models for the integrated light from galxies, van
Dokkum \& Conroy (2010) showed that the Wing-Ford band in the centers
of giant elliptical galaxies is much stronger than expected from
standard simple stellar populations models.  Later, in Conroy \& van
Dokkum (2011), they argued further that the much smaller elliptical
galaxy NGC~4458 had an initial mass function much closer to the
Salpeter IMF, again on the basis of its Wing-Ford band.  An
alternative method for testing the initial mass function has also been
proposed recently, and has also found a higher M/L ratio in the most
massive galaxies than is expected from a Salpeter, Kroupa, or Chabrier
IMF (Cappellari et al. 2012), but this work examines {\it only} the
stellar mass to light ratio, and can give the same results for a
bottom-heavy IMF dominated by dwarf stars, or a top heavy IMF with
larger numbers of black holes, neutron stars and white dwarfs.

\begin{figure}
\includegraphics[angle=-90,width=9 cm]{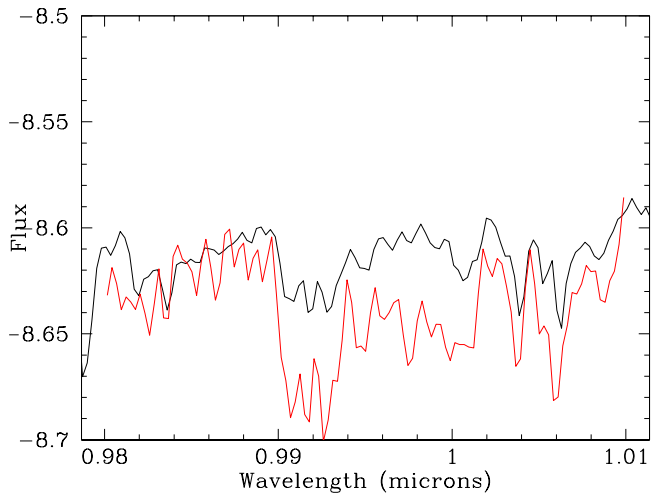}
\caption{The spectra of two similar temperature stars from the IRTF
  library (Rayner et al. 2009).  The upper curve, in black, is the
  spectrum of HD18191, an M6III star.  The lower curve, in red, is the
  spectrum of SU Mon, a S6 star.  The continua shortward of the
  Wing-Ford band are similar, while the spectrum along the
  Wing-Ford band shows suppressed flux for the S star,
  indicating that its Wing-Ford band strength is much larger.  Some S
  stars show even deeper Wing-Ford bands, but these are not in the
  IRTF libraries.}
\label{SU Mon}
\end{figure}

It has also been found, however, that stars rich in s-process elements
show strong Wing-Ford bands (Wing 1972; Lambert \& Clegg 1980),
leading to a flux reduction of about 0.1-0.2 magnitudes.  This can
also be seen from the IRTF spectra of S stars (Rayner et al. 2009 --
see also figure \ref{SU Mon}).  Since the Wing-Ford feature in the
giant elliptical galaxies is only about 0.02 magnitudes deeper than
expected from a Salpeter initial mass function, only about 10\% of the
giants' light needs to come from S stars for the S stars to explain
the deviation from the predictions of a Salpeter IMF.  Since the
Wing-Ford band itself is quite broad with the deep part of the
absorption spanning about 40 \AA, and the whole band spanning more
than 100\AA, the smearing due to the few hundred km/sec velocity
dispersions in elliptical galaxies will be negligible -- unlike the
case for narrow spectral features which might be strongly affected.

It should be noted that the empirical spectral libraries used by van
Dokkum and Conroy (2010) explicitly excluded giants of unusual
chemical composition, rather than attempting to estimate the number of
such stars and weigh their empirical spectra accordingly.  As a
result, their models are missing single-star channel S stars for all
galaxies, and is also missing the binary channel S stars for galaxies
where they can exist.  The single star channels are not likely to be
especially important for early type galaxies, since the s-process
elements reach the surface only during the third dredge up phase of
stellar evolution, a process which happens only for relatively high
mass stars.

In order to estimate the magnitude of the effect on the Wing-Ford
band, we need an estimate of the fraction of giants which are S stars.
This is not straightforward to do from the literature, as the number
of very cool giants is rather small, and in the CCD era, only recently
has there been good enough sensitivity at 9900 \AA for spectra at that
wavelength to be common.  As one manner of estimating the number of S
stars, we can rely on the number densities of barium stars, which are
identified by features around 4400 \AA, and which are thought to be
progenitors of the extrinsic S stars, and we can assume that the ratio
of S stars to M giants will be similar to the ratio of barium stars to
earlier type red giants.  We also note that roughly half of the S
stars seen in the Milky Way in flux limited surveys are intrinsic and
roughly half are extrinsic (e.g. Yang et al. 2006).

Following a suggestion by Han et al. (1995), we restrict the counts of
the two classes of objects to those stars brighter than 6th magnitude,
so the stars will be members of the Bright Star Catalog (Hoffleit \&
Warren 1995), and also allows us to be reasonably confident that most
of the barium stars will have been identified as such.  We also note
that the barium stars have been found to show typically the same
absolute magnitudes as other G/K giants (e.g. Kemper 1975; Hakkila
1990).  Restricting ourselves to giants with spectral classes from G5
to K5, we find 61/1194, or 5.5\% of the stars are barium stars -- we
regard this number as a lower limit, since there does not exist a
definitive paper with upper limits that definitively show that all the
barium stars are accounted for.  This limit is considerably higher
than what was estimated through a similar procedure by Warner (1965)
-- however Warner noted that the barium stars in his sample already
skewed toward being around K0 in spectral type, and that relatively
fewer normal giants were in that temperature range than is the case
for later K-types.

Similarly, about 4\% of the M giants in the Bright Star Catalog
(Hofleit \& Warren 1995) correspond with S stars in the catalog of
Stephenson (1984).  Keenan (1954) found that roughly 10\% of M giants
were S stars as well, and no more recent systematic attempt seems to
have been made to estimate the fraction of stars occupying the M-giant
region of the color magnitude diagram which are S stars.  CH stars
account for roughly 30\% of halo giants (e.g. Lucatello et al. 2005).

We can thus conclude that removal of the extrinsic S stars from a
galaxy will have a significant effect on its apparent initial mass
function if the mass function is measured using the Wing-Ford band.
If $\sim$ half of the barium stars are as yet unidentified, then the
extrinsic S stars may explain the entire effect on the Wing-Ford band,
with no need for an altered stellar initial mass function.  Another
issue, whose effect is not clear, is whether the binary fractions of
galaxies vary systematically.  Within the Galaxy, there is some
suggestive evidence that metal rich stars have a higher binary
fraction than metal poor stars (e.g. Riaz et al. 2008), which might
also boost the number of S stars in massive galaxies relatively to
that in lower mass galaxies.  Studies have been done of extragalactic
binary fractions only in a few very nearby galaxies.  They are within
a factor of the few of the disk binary fraction, but with large
statistical uncertainties (e.g. Geha et al. 2013).

One clear prediction of a scenario where the S stars are significant
contributors to the Wing-Ford band fluxes of galaxies is that the ZrO
bands should be stronger in such galaxies as well.  In the most
comprhensive analysis to date of elliptical galaxies (van Dokkum \&
Conroy 2012), there is no real sensitivity to these bands.  In
integrated light, the strong bands near 4600 \AA will be swamped out
by the light from turnoff stars, while the strong band around 6470 \AA
is not covered in the spectra, and the region around the strong
infrared band of ZrO at about 9300 \AA is in the region of strong sky
background.

We note that the Na~I~D absorption line was also presented by van Dokkum \&
Conroy (2010) as part of the evidence for a bottom heavy initial mass function
in giant elliptical galaxies.  This line is also strongly enhanced in barium
stars and S stars -- Warner (1965) estimates that the sodium abundances of the
barium stars are roughly 1.2-1.8 times larger than expected for stars of the
same iron abundance.  Additionally, the [Na/Fe] abundance ratio increases with
increasing [Fe/H], at least for super-solar stars in the Milky Way disk
(e.g. Edvardsson et al. 1993).  The use of solar metallicity template spectra
will therefore affect sodium absorption lines much more strongly than most
other lines.

An alternative test of the scenario is to look at the central region
of the Milky Way.  Dynamical encounters there have been suggested to
affect the formation rates of X-ray binaries (Maccarone \& Patruno
2013; see also a similar suggestion for the inner bulge of M31 from
Voss \& Gilfanov 2007).  In the central regions of the Milky Way, the
stellar density can be $\sim10^6$ stars per pc$^3$ (i.e. in the
nuclear star cluster -- Genzel et al. 2010).  The density remains
greater than a few thousand solar masses per cubic parsec out to
several parsecs.  Since the absolute $K$ band magnitudes of the
extrinsic S stars are about -5.7 (van Eck et al. 1998), they should
have apparent magnitudes of roughly 14 at the distance of the Galactic
Center, assuming a distance of 8 kpc, and 5 magnitudes of extinction
-- these sources should be detectable as S stars in surveys of the
Galactic Center, e.g. with FLAMINGOS-2 (Eikenberry 2008).

\section{Conclusions}
Binary destruction in the cores of the smaller giant elliptical
galaxies is expected.  A lack of S stars in these galaxies can provide
an alternative explanation for the recent claims of steeper initial
mass functions in the biggest giant elliptical galaxies than in
smaller ellipticals, eliminating the need to deal with the conflict
between those results and other estimates of the dark matter content
of the Universe.  This idea can be tested by searching for the ZrO
bands in giant elliptical galaxies, and by spectroscopic follow-up of
giants in the center of the Milky Way relative to giants elsewhere in
the Milky Way.

If the Wing Ford band can be confirmed to be dominated, or even
affected, by the S stars, the results from Cappellari et al. (2012)
showing higher M/L in higher velocity dispersion galaxies remain
unaffected by the conclusions of this work.  However, the results of
Cappellari et al. (2012) have some degeneracies between the functional
form of the dark matter density distribution and the stellar
mass-to-light ratio.  They can also produce higher-than-standard M/L
ratios either through top heavy or bottom heavy initial mass
functions, with the former producing high M/L ratios due to having
more compact objects, and the latter due to having more low luminosity
M-dwarfs.  Understanding all the systematic effects in stellar
population models is thus a key for understanding the root cause of
the results of Cappellari et al (2002).

\section{Acknowledgments}
I thank Anthony Gonzalez, Steve Zepf, Claudia Maraston and Dave Sand for useful
discussions.

\label{lastpage}


\begin{thebibliography}{99}
\bibitem{}Belczy\'nski K., Miklajewska J., Munari U., Ivison R.J., Friedjung
  M., 2000, A\&AS, 146, 407 
\bibitem{}Belczynski K., Kalogera V., Rasio F.A., Taam R.E., Zezas A., Bulik T., Maccarone T.J., Ivanova N., 2008, ApJS, 172, 233
\bibitem{}Bidelmam W.P., Keenan P.C., 1951, ApJ, 114, 473
\bibitem{}Binney J., Tremaine S., 2009, ``Galactic Dynamics'' Princeton University Press: Princeton, NJ, USA
\bibitem{}Dressler A., Lynden-Bell D., Burstein D., Davies R.L., Faber S.M., Terlevich R., Wegner G., 1987, ApJ, 313, 42
\bibitem{}Edvardsson B., Andersen J., Gustafsson B., Lambert D.L., Nissen
  P.E.,  Tomkin J., 1993, A\&A, 275, 101
\bibitem{}Eikenberry S., 2008, AIPC, 1010, 132
\bibitem{}Fregeau J., 2008, ApJ, 673L, 25
\bibitem{}Gebhardt K., Thomas J., 2009, ApJ, 700, 1690
\bibitem{}Gebhardt K., et al., 1996, AJ, 112, 105
\bibitem{}Geha M., et al., 2013, ApJ, 771, 29
\bibitem{}Genzel R., Eisenhauer F., Gillessen S., 2010, RevMP, 82, 3121 
\bibitem{}Hakkila J., 1990, AJ, 100, 2021
\bibitem{}Han Z., Eggleton P.P., Podsiadlowski P., Tout C.A., 1995, MNRAS, 277, 1443
\bibitem{}Han Z., Podsiadlowski P., Maxted P.F.L., Marsh T.R., Ivanonva N., 2002, MNRAS, 336, 449
\bibitem{}Han Z., Podsiadlowski P., Lynas-Gray A.E., 2007, MNRAS, 380, 1098
\bibitem{}Hardy E., Couture J., 1988, ApJ, 325L, 29
\bibitem{}Heggie D.C., 1975, MNRAS, 173, 729
\bibitem{}Hofleit D., Warren W.H., 1995, ``Bright Star Catalog'', 5th edition
\bibitem{}Iben I., Renzini A., 1983, ARA\&A, 21, 271
\bibitem{}Ivanova N., 2012, ApJ, 760L, 24
\bibitem{}Jorissen A., Mayor M., 1992, in ``Evolutionary Processes in Interacting Binary Stars.'' Proceedings of the 151st Symposium of the International Astronomical Union, editors, Y. Kondo, R. F. Sistero, R. S. Polidan; Kluwer Academic Publishers: Dordrecht
\bibitem{}Keenan P.C., 1942, ApJ, 96, 101
\bibitem{}Kemper E.C., 1975, PASP, 87, 537
\bibitem{}Lambert D.L., Clegg R.E.S., 1980, MNRAS, 191, 367
\bibitem{}Lucatello S., Gratton R.G., Carretta E., Beers T.C., Christlieb N., Cohen J.G., HiA, 13, 587
\bibitem{}Maccarone T.J., Patruno A., 2013, MNRAS, 428, 1335
\bibitem{}Merrill P.W., 1922, ApJ, 56, 475
\bibitem{}McClure R.D., 1985, JRASC, 79, 277
\bibitem{}McClure R.D., Woodsworth A.W., 1990, ApJ, 352, 709
\bibitem{}Rayner J.T., Cushing M.C., Vacca W.D., 2009, ApJS, 185, 289
\bibitem{}Riaz B., Gizis J.E., Samaddar D., 2008, ApJ, 672, 1153
\bibitem{}Schaefer B., 2014, AAS, 223, 335.04
\bibitem{}Sokoloski J., 2003, JAAVSO, 31, 89
\bibitem{}Stephenson C.B., 1984, Publ. Warner \& Swasey Obseravtory, 3, 1
\bibitem{}Sugimoto D., Bettwieser E., 1983, MNRAS, 204, 19
\bibitem{}van Dokkum P.G., Conroy C., 2010, Nature, 468, 940
\bibitem{}van Dokkum P.G., Conroy C., 2011, ApJ, 735L, 13
\bibitem{}Van Eck S., Jorissen A., Udry S., Mayor M., Pernier B., 1998, A\&A
  329, 971 
\bibitem{}Voss R., Gilfanov M., 2007, A\&A, 468, 49
\bibitem{}Warner B., 1965, MNRAS, 129, 263
\bibitem{}Whitford A.E., 1977, ApJ, 211, 527
\bibitem{}Wing R.F., Ford W.K., 1969, PASP, 81, 527
\bibitem{}Wing R.F., 1972, Les spectres des astres dans l'infrarouge et les microondes, p. 123
\bibitem{}Yang X., Chen P., Wang J., He J., 2006, ApJ, 132, 1468
\bibitem{}Yoo J., Chanam\'e J., Gould A., 2004, ApJ, 601, 311

\end{thebibliography}
\end{document}